# Navigating Ethical Challenges in Generative AI-Enhanced Research: The ETHICAL Framework for Responsible Generative AI Use

**Douglas Eacersall**, Library Services, University of Southern Queensland, Australia and University of the Philippines, Open University, Philippines. ORCID: https://orcid.org/0000-0002-2674-1240

**Lynette Pretorius**, School of Curriculum, Teaching, and Inclusive Education, Faculty of Education, Monash University, Melbourne, Australia. ORCID: https://orcid.org/0000-0002-8998-7686

**Ivan Smirnov**, Graduate Research School, University of Technology Sydney. ORCID: https://orcid.org/0000-0002-8347-6703

**Erika Spray**, School of Education, University of Newcastle, Australia. ORCID: https://orcid.org/0000-0002-0367-4508

**Sam Illingworth**, Department for Learning and Teaching Enhancement, Edinburgh Napier University, Edinburgh, UK. ORCID: https://orcid.org/0000-0003-2551-0675

**Ritesh Chugh**, School of Engineering & Technology, CML-NET Research Centre Central Queensland University, Australia. ORCID: https://orcid.org/0000-0003-0061-7206

**Sonja Strydom**, Centre for Learning Technologies, Stellenbosch University, South Africa. ORCID: https://orcid.org/0000-0003-2270-2209

**Dianne Stratton-Maher**, School of Nursing and Midwifery. University of Southern Queensland. ORCID: https://orcid.org/0000-0001-9191-5588

**Jonathan Simmons**, Writing Services, University of Alberta, Canada. ORCID: https://orcid.org/0000-0001-5231-0578

**Isaac Jennings**, eResearch and Specialised Platforms, Griffith University. ORCID: https://orcid.org/0009-0001-4723-4045

**Rian Roux**, UniSQ College, University of Southern Queensland. ORCID: https://orcid.org/0000-0003-0278-3195

**Ruth Kamrowski**, Griffith Graduate Research School, Griffith University. ORCID: https://orcid.org/0009-0009-7947-1684

**Abigail Downie**, Griffith Graduate Research School, Griffith University. ORCID: https://orcid.org/0009-0005-1528-2789

**Chee Ling Thong**, Institute of Computer Science and Digital Innovation, UCSI University, Malaysia. ORCID: https://orcid.org/0000-0002-5138-883X

**Katharine A. Howell**, Graduate Research, Edith Cowan University, Perth, Australia. ORCID: https://orcid.org/0000-0002-5044-0034

*Corresponding author: Douglas.Eacersall@unisq.edu.au

# Abstract

The rapid adoption of generative artificial intelligence (GenAI) in research presents both opportunities and ethical challenges that should be carefully navigated. Although GenAI tools can enhance research efficiency through automation of tasks such as literature review and data analysis, their use raises concerns about aspects such as data accuracy, privacy, bias, and research integrity. This paper develops the ETHICAL framework, which is a practical guide for responsible GenAI use in research. Employing a constructivist case study examining multiple GenAI tools in real research contexts, the framework consists of seven key principles: 'Examine policies and guidelines', 'Think about social impacts', 'Harness understanding of the technology', 'Indicate use', 'Critically engage with outputs', 'Access secure versions', and 'Look at user agreements'. Applying these principles will enable researchers to uphold research integrity while leveraging GenAI's benefits. The framework addresses a critical gap between awareness of ethical issues and practical action steps, providing researchers with concrete guidance for ethical GenAI integration. This work has implications for research practice, institutional policy development, and the broader academic community while adapting to an AI-enhanced research landscape. The ETHICAL framework can serve as a foundation for developing AI literacy in academic settings and promoting responsible innovation in research methodologies.

## Introduction

The term Artificial Intelligence (AI) has existed for at least 60 years (Cordeschi, 2007). Since the recent emergence of generative AI (GenAI) tools, such as ChatGPT, there has been an increased focus on AI within academia and the mainstream media. GenAI uses machine learning, incorporating large amounts of training data, algorithms and models to create text, images, code and audio (Bandi et al., 2023). Given the potential of GenAI to perform tasks currently only undertaken by humans, the implications for various professions and industries, including research, are significant. GenAI's influence is transforming traditional research methodologies, inviting both enthusiasm and scrutiny regarding its role in academic integrity and the quality of scholarly work. This dual impact underscores the need for clear ethical guidelines and new training protocols to integrate GenAI effectively in research settings.

In recent years, numerous GenAI tools have been developed, offering great potential to enhance various aspects of the research process. These tools have shown promise across multiple areas, including research idea generation (Dowling & Lucey, 2023; Pretorius, 2023; Pretorius et al., 2024; Xu et al., 2023), survey and questionnaire design (Götz et al., 2024), and literature search (Alqahtani et al., 2023; Extance, 2018). GenAI also aids in data generation (Pretorius & Cahusac de Caux, 2024), annotation (Matter et al., 2024), and augmentation (Kim & Lee, 2024), providing researchers with advanced tools to improve data quality and relevance. Additionally, GenAI models enable the simulation of human behaviour (Argyle et al., 2023; Aher et al., 2023), which can be beneficial in social science research. They can also be used to assist with complex data analysis tasks (Alqahtani et al., 2023; Christou, 2023; Pretorius & Cahusac de Caux, 2024). Moreover, these technologies have applications in academic writing (Lucey & Dowling, 2023; Lund et al., 2023; Mojadeddi & Rosenberg, 2023; Orenstein, 2023; Pretorius et al., 2024), science communication (Alvarez et al., 2024), peer review and publishing (Razack et al., 2021), and even research funding processes (Spyroglou et al., 2022).

GenAI's potential to revolutionise research practices lies in its capacity to automate traditionally time-consuming tasks, such as conducting literature reviews, cleaning datasets, and performing preliminary analyses. This automation can increase research efficiency and contribute to the quality of research outcomes, potentially improving the overall output and well-being of researchers (Bail, 2024; Sawyer & Henriksen, 2024). Furthermore, by facilitating real-time translation and interpretation, GenAI can bridge linguistic and cultural gaps, fostering more inclusive collaboration across international borders (Baldassarre et al., 2023; Pretorius et al., 2024). Through these diverse applications, GenAI stands poised to reshape the research landscape, offering tools that not only support existing research practices but also open new possibilities for innovation and cross-cultural exchange.

Although GenAI-assisted research tools have great potential to aid the research endeavour, many challenges have been identified with their use. These have mainly been experienced in relation to research integrity and social impact.

In terms of research integrity, the use of GenAI tools raises several issues, particularly regarding data accuracy, security, reliability, and authorship. For example, the nature of GenAI means that its use may jeopardise a researcher's ability to meet privacy expectations required for ethics clearance or to protect commercial-in-confidence information. Another major issue is the propensity for these tools to produce factually incorrect content, known as hallucinations (Mahyoob et al., 2023; Manakul et al., 2023), meaning that data and text produced can be factually incorrect or lacking in nuance. Transparency is also a major concern (Dwivedi et al., 2023), as information regarding how the GenAI tool was trained is often not available. This is referred to as the black box nature of AI (Barr, 2023; Bearman & Ajjawi, 2023) and makes it difficult for researchers to ascertain the reliability and recency of the data used to produce outputs, with consequences for the replicability of results. As GenAI tools construct new material from training data based on existing works, ethical concerns have been raised regarding authorship, copyright and plagiarism (Frye & ChatGPT, 2023; Lund et al., 2023). For instance, tools such as ChatGPT may incorrectly reference the text created (Lund et al., 2023; Walters & Wilder, 2023), making it difficult to accurately acknowledge other authors' work. Ongoing lawsuits related to GenAI applications using copyrighted material without permission during model training (Franceschelli & Musolesi, 2022) underscore these concerns. Some AI models may have been trained on copyrighted works, and if researchers use outputs that closely resemble those works, legal disputes could arise.

It is also important to consider the social impact of GenAI use in research. This mainly involves the perpetuation of bias, equity considerations, and environmental impact. Generative AI outputs are prone to bias, reflecting the political (Hartmann et al., 2023), racial (Zhang et al., 2023), disability (Hutchinson et al., 2020) and gender prejudices (Basta et al., 2019) contained in the large datasets on which they have been trained. Equitable access is another issue in this space. Much of the debate is situated within wealthy, developed economies where a large proportion of the population has access to GenAI. This is not necessarily true globally and may, therefore, reinforce the disadvantage of researchers working in less privileged contexts. There is also a growing concern regarding the environmental costs associated with GenAI (Bender et al., 2021), as well as questionable work practices by dominant GenAI firms (Tan & Cabato, 2023).

Compounding each of the issues outlined above is the pace of change in AI technologies. GenAI tools are emerging, undergoing further development, and being embraced by users at such a rapid rate that it is difficult for stakeholders to keep pace and adapt accordingly (Boyd & Wilson, 2017; Mingsakul, 2023).

GenAI literacies need to be developed to enhance the benefits of GenAI and mitigate the challenges. The AI Literacy: Principles of ETHICAL Generative Artificial Intelligence resource (Eacersall, 2024) was developed as a starting point to address this issue. In this current article, we expand on these principles and present an ethical framework to address the research question: *What practical knowledge and guidance would assist researchers in effectively navigating the ethical challenges in generative AI-enhanced research?*

Much of the current literature discusses the challenges of ethical GenAI-enhanced research, but there is limited guidance on the appropriate actions that researchers should take. Therefore, the focus of this research is not only on identifying ethical issues, but also on what researchers can practically do to ethically navigate the GenAI-enhanced research space. In discussing the considerations for researchers and what they should do, this research also informs regulators, research institutions and GenAI developers so that they can better support and guide researchers and develop GenAI-enhanced research tools in ways that limit risk and maximise benefits.

## Methodology

The ETHICAL framework was developed using a constructivist single case study approach that examined the use of GenAI tools in research practice. This involved the first author (DE) using multiple GenAI tools to assist with project conceptualisation, literature review and analysis, and grant writing in the context of a real research project. The AI outputs and research processes for each tool were recorded and then compared against each other, as well as more traditional research approaches. The first author (DE) also maintained a reflective research journal to record relevant insights into the use of GenAI tools for these aspects of the research process. These reflections were then analysed using a reflexive thematic analysis approach (Braun & Clarke, 2006, 2019, 2022). This involved reading and re-reading the text, inductive coding to highlight key points, and the construction of themes using the researcher's reflexivity as a lens (Braun & Clarke, 2022). Through this process, seven initial themes were identified and named to align with the acronym ETHICAL. These initial themes were shared at several conferences and workshops via an institutional research repository (see Eacersall, 2024), and feedback was gathered to further inform the process.

At this point, DE worked with LP, who acted as a critical friend to further refine the themes so that they presented distinct ideas. To then construct a framework, DE contacted a group of other global educators and researchers involved in AI research and practice. All of these authors worked together collaboratively to develop detailed descriptions of each theme and actionable steps for others to use. In this way, the trustworthiness of this qualitative approach was established through researcher reflexivity and agreement among co-researchers.

## Findings and Discussion

The seven themes identified through the methodological process described above were: **E** - Examine policies and guidelines, **T** - Think about the social impacts, **H** - Harness understanding of the technology, **I** - Indicate use, **C** - Critically engage with outputs, **A** - Access secure versions, and **L** - Look at user agreements. These seven themes were joined together to create the ETHICAL Framework for Responsible Generative AI Use, which is displayed in Figure 1. The framework illustrates that the application of these ethical principles is not a one-time action but rather an ongoing, reflective process. This is emphasised using circular arrows, which symbolise the continuous, cyclical nature of ethical decision-making. As individuals engage with these principles, they are encouraged to revisit and reassess their actions, considering new information, evolving contexts, and shifting perspectives. Each component of the framework is discussed in detail below.

**Figure 1**

*The ETHICAL Framework for Responsible Generative AI Use*

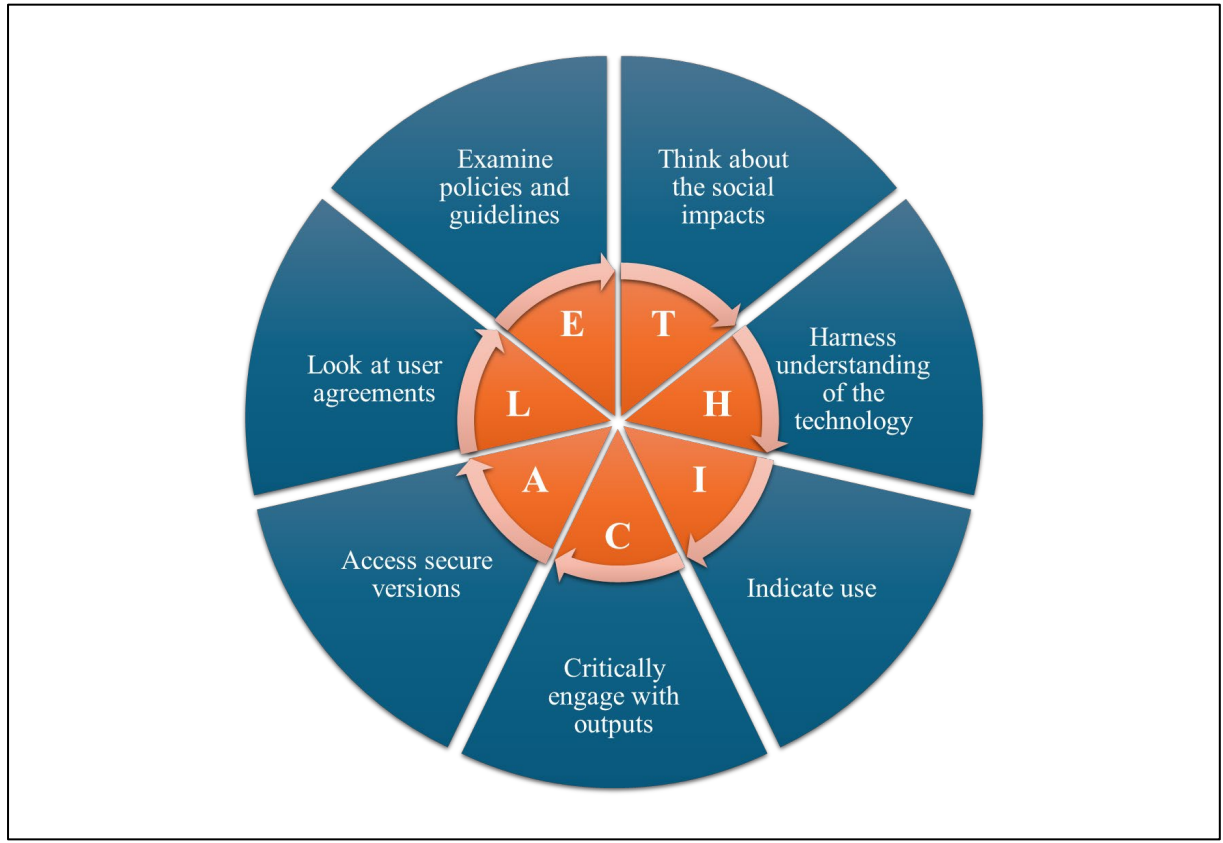

## E - Examine policies and guidelines

When examining policies and guidelines for GenAI, it is crucial to carefully review those relevant to the researcher's specific context, starting with broad international regulations, such as the European Commission's *Living guidelines on the responsible use of generative AI in research* and then narrowing down to specific national and local standards to ensure compliant and ethical use. These international frameworks guide the development, deployment, and use of AI technologies, focusing on trust, safety, and ethical standards (European Commission, 2024; Novelli et al., 2024), establishing benchmarks that influence global practices and ensure compliance across borders. In addition to government regulations, ethical AI frameworks developed by non-governmental bodies, such as UNESCO's AI ethics recommendations, also play a critical role in shaping global standards for responsible AI use (UNESCO, 2023). These non-state actors provide additional guidance that complements formal legislation.

At the national level, it is important to consider country-specific regulations that may adapt or expand international guidelines to address local legal and ethical issues. Different nations may have unique requirements reflecting their legal environments and societal values. For instance, Australia's Tertiary Education Quality and Standards Agency (TEQSA) provides guidelines for the responsible use of AI in research in Australia (TEQSA, 2024). Additionally, Australia's policy for the responsible use of AI in government aims to ensure that AI is adopted with confidence and ethical integrity, enhancing public trust through transparency, governance, and an approach that aligns with community expectations in Australia (Australian Government, 2024).

Locally, policies may further tailor these broader guidelines to address regional practices and institutional norms. For researchers, this means understanding how local data protection laws and intellectual property regulations apply to their use of GenAI tools, ensuring compliance with national and international standards. Additionally, individuals, including researchers, must be aware of intellectual property rights related to content generated by GenAI to avoid potential legal disputes and respect creative rights (Smits & Borghius, 2022). Sector-specific policies may also influence how GenAI is regulated and applied in different industries, such as healthcare, finance, or higher education. Organisations in these sectors may need to adapt GenAI use based on specific ethical or operational concerns. Becoming acquainted with sector-specific and internal organisational policies will enable researchers to align their GenAI practices more effectively with the expectations and demands of their workplace setting (Dwivedi et al., 2021).

Integrating international best practices with national and local regulations helps organisations, including research institutions, develop robust, context-specific policies addressing data privacy, transparency, accountability, and intellectual property issues (Lescrauwaet et al., 2022). Implementing this approach ensures that AI systems are legally compliant and ethically sound within their jurisdictions. Additionally, organisations must provide training and resources to their staff to foster an understanding of these layered regulatory requirements (Kutty et al., 2024).

## T - Think about the social impacts

The integration of GenAI into research necessitates a critical examination of its broader societal ramifications. These include bias, equity, and environmental sustainability considerations.

While GenAI systems offer powerful capabilities, their reliance on extensive datasets can perpetuate existing biases, potentially leading to discriminatory outcomes (Ferrara, 2023). These biases can manifest in the reinforcement of harmful stereotypes and introduce unfairness into decision-making processes within critical domains such as healthcare and education (Meskó & Topol, 2023; Farrelly & Baker, 2023; Pretorius & Cahusac de Caux, 2024). The inherent political and social ideologies embedded within AI models pose an additional challenge, as these tools may inadvertently propagate specific worldviews (Christodoulou & Iordanou, 2021).

The increasing adoption of GenAI in research also accentuates the persistent issue of the digital divide. Unequal access to advanced AI tools creates a distinct disadvantage for researchers lacking high-performance computing infrastructure or specialised expertise. This disparity limits the participation of underrepresented groups and threatens to solidify existing inequalities within academia. The concentration of research power in well-resourced institutions or countries is a potential consequence (Lutz, 2019; Kitsara, 2022). Additionally, the cost and infrastructure requirements of GenAI can marginalise researchers in less affluent regions (Spyroglou et al., 2022).

The rapid advancement and deployment of GenAI systems raises significant concerns about their environmental sustainability. The computational power required to train and run large AI models results in substantial energy consumption and associated carbon emissions. These environmental costs are often overlooked, and therefore, more energy-efficient AI architectures and deployment strategies are required. Researchers should be cognisant of these costs and consider the trade-offs between model performance, energy consumption, and the suitability of tools for specific research use. Environmental impact adds another layer of complexity to the ethical considerations surrounding GenAI, necessitating a holistic approach that balances technological advancement with ecological responsibility (Luccioni et al., 2024).

Despite the challenges, the responsible application of GenAI offers significant societal benefits. Novel AI tools are empowering neurodivergent individuals by enhancing communication and reducing misunderstandings (Arango et al., 2024). For example, AI-powered writing assistants can aid in understanding and adjusting the tone of written communication, promoting clarity and minimising misinterpretations. GenAI can also develop personalised learning tools that cater to the unique needs of neurodivergent individuals, fostering a more inclusive and accessible learning environment.

AI-powered tools, such as real-time captioning, transcription, and sign language interpretation, can significantly enhance accessibility for individuals with disabilities. These technologies facilitate greater participation in various aspects of society, including education and the workforce (Chemnad & Othman, 2024; ZainEldin et al., 2024). GenAI's capacity to create personalised learning experiences can be particularly transformative, tailoring educational content and delivery methods to individual needs and learning preferences (Kestin et al., 2024). This approach democratises access to information and empowers learners of all ages and stages to pursue their interests and achieve their full potential.

While concerns about GenAI's negative impacts persist, it is important to recognise its potential to reshape the landscape of opportunities. By addressing language barriers, improving digital literacy, automating routine tasks, and providing cost-effective coaching, GenAI can contribute to economic growth and social development. For example, in terms of employment, while some jobs may be displaced, GenAI also has the potential to create new types of jobs and augment existing ones, particularly in developing countries (Mannuru et al., 2023).

Although acknowledging the potential negative impacts of GenAI is paramount, these challenges are not insurmountable. Proactive and responsible development and deployment practices, coupled with investments in upskilling and reskilling programs, are necessary for harnessing GenAI's transformative power while mitigating its potential downsides (Morandini et al., 2023). GenAI's inherent capacity to remedy its own negative social impacts is often underemphasised. Advancements in AI-powered privacy-enhancing technologies can help safeguard sensitive data and protect individual privacy (Abolaji & Akinwande, 2024; Muhammad et al., 2023; Yanamala, 2023). By automating routine tasks, GenAI can free up human workers to focus on more creative and strategic endeavours, fostering a more fulfilling and productive work environment (Haase & Hanel, 2023).

The deployment of GenAI in research necessitates a balanced approach, acknowledging both its potential to perpetuate social inequalities and its ability to contribute positively to research practices. In engaging with these tools, researchers need to critically consider the societal impacts and ensure that the use of GenAI in research promotes inclusivity and equity and has minimal negative impact on society.

## H - Harness understanding of the technology

In a world increasingly driven by AI technology, it is crucial that researchers understand how specific AI systems operate, including the capabilities and limitations of individual tools. Such knowledge is not merely beneficial but essential, as it plays a critical role in the ability to evaluate the information generated by these systems (Solomon & Davis, 2023). Through understanding the underlying mechanisms and algorithms of AI systems, individuals develop effective AI literacy (Markus et al., 2024). AI literacy encompasses a range of skills that enable individuals to critically assess AI technologies, interact with AI systems, and use AI effectively as a tool (Long & Magerko, 2020; Pretorius & Cahusac de Caux, 2024). This involves not only basic usage but also the critical evaluation of AI outputs to assess their quality and accuracy (Almatrafi et al., 2024; Pretorius & Cahusac de Caux, 2024). Ultimately, mastering these skills ensures responsible and informed engagement with AI technology.

Furthermore, harnessing an understanding of how AI systems function is essential for evaluating their ethical implications (Bankins & Formosa, 2023). Grasping the underlying mechanisms and algorithms facilitates the identification of potential ethical considerations inherent in the generated content. Such knowledge ensures that AI outputs adhere to established ethical standards and do not inadvertently result in negative consequences (Lund et al., 2023). Additionally, a thorough understanding of AI systems promotes a more responsible and informed use of these technologies, guiding researchers toward ethical and effective practices.

By harnessing this technological understanding, researchers can more effectively navigate and utilise AI systems to their fullest potential. This knowledge empowers them to make informed decisions about AI deployment and to maximise the technology's benefits while remaining mindful of its limitations and challenges. Ultimately, developing a robust understanding of AI allows researchers to manage the complexities of this technology more effectively, ensuring that its application is both responsible and beneficial.

## I - Indicate use

It is imperative that researchers clearly indicate how GenAI has been utilised in their work. Research suggests that tools that currently purport to detect AI-generated or adapted text have significant limitations regarding accuracy and reliability (Weber-Wulff et al., 2023). This calls for transparency by authors and is crucial for several reasons (Crawford et al., 2023). Firstly, it addresses issues of authorship by distinctly differentiating between human-generated and AI-generated content, thereby ensuring that credit is appropriately allocated. This clarity is essential in academic and professional contexts where the integrity of authorship is paramount (Perkins, 2023). Secondly, transparency regarding AI usage can mitigate potential copyright issues by ensuring that all sources and contributions are properly acknowledged (Bukhari et al., 2023). Finally, openly disclosing the use of AI fosters trust and credibility, both within the researcher's professional community and with the broader public. It demonstrates a commitment to ethical standards and accountability, indicating an awareness of the implications and responsibilities associated with employing advanced technologies. By explicitly stating the role of AI in their work, researchers contribute to a culture of openness and ethical rigour.

There are some practical guidelines that could assist researchers in indicating GenAI usage. For instance, disclosures with clear documentation of the level of GenAI usage positively contribute to transparency and academic integrity (Hosseini et al., 2023). Furthermore, a complete grasp of institutional and/or publisher guidelines assists in addressing copyright issues and the "essence of academic discourse" (Carobene et al., 2024, p. 835). In addition, authors are encouraged to ensure a working understanding of ethical issues and conundrums associated with GenAI usage (Bearman et al., 2022). There is also guidance for citing GenAI in referencing guidelines such as those provided by the American Psychological Association.

Although there are already guidelines to help researchers outline the use of GeanAI in their work, we suggest that more guidance is needed in this space. Researchers need to develop and codify both general and discipline-specific GenAI methodologies, encompassing all aspects of the research process (e.g. project conceptualisation, literature review and organisation, data collection and analysis, and write-up and dissemination). These methodologies need to incorporate systematic processes for documenting ethical GenAI use in terms of what tools were used, how they were used, and to what extent, as well as outlining adherence to relevant ethical guidelines and/or frameworks.

## C - Critically engage with outputs

The responsibility for the accuracy, ethics and standards of disseminated research outputs ultimately rests with the authoring researcher, not the technology platform. GenAI technologies, whilst powerful, are not infallible. The defacto understanding of truth that AI models operate on is known as "ground truth" or "fundamental truth" and this (human) construction effectively becomes the "reality" against which developers measure the model (Munn et al., 2023, pp. 2-3). In this regard, system designers who select the limited data that the system is trained on and who provide operational parameters ultimately hold the power to determine the kind of outputs that are generated or not (Crawford, 2022; Munn et al., 2023). This means that AI models may generate inaccurate, biased or inappropriate outputs, and researchers must, therefore, critically engage with and revise all content generated by artificial intelligence. Critical engagement entails being mindful of how knowledge is acquired and justified within specific disciplines and how risks can be proactively mitigated. Criticality in this regard is what allows a researcher to leverage the unique proficiencies of GenAI tools whilst maintaining epistemic responsibility.

By critically engaging with and revising outputs, the researcher can ensure that their contribution to knowledge is reliable and valid. This necessarily entails making informed judgements regarding the truth value or accuracy of generated content in relation to their specific discipline areas. It involves actively checking that artificially generated outputs both correspond well to the available evidence and are coherent with mutually implicative information or ideas (Roux, 2024). This process ensures that GenAI tools inform and enhance the researcher's work, rather than supplanting their intellectual, ethical and creative processes. In this manner, the researcher retains control over the quality and direction of their work, ensuring that their academic and professional expertise and judgement are utilised effectively to complete the final product.

## A - Access secure and protected versions

Although GenAI tools have enormous potential to transform efficiency in research, they
utilise vast amounts of data, which can result in serious risks to privacy and security (Chen &
Esmaeilzadeh, 2024). It is important for researchers using GenAI to evaluate and assess the
security of their chosen GenAI tool(s), considering copyright, intellectual property, privacy,
legal, ethical and research integrity responsibilities, to ensure the integrity of their work and
avoid issues related to plagiarism and unauthorised use of proprietary information (Golda et
al., 2024).

GenAI tools learn from curated internet data sets (this may include publicly available,
copyrighted and pirated materials), user inputs, previous outputs, and feedback from users
(O'Leary, 2013; Fui-Hoon et al., 2023). Literature provides a variety of cautionary tales
associated with data leakage across various settings, ranging from sensitive medical data to
accidental disclosure of confidential company information and attackers entering fake data to
undermine software and systems developed using GenAI (Chen & Esmaeilzadeh, 2024;
Gupta et al., 2023; Khanna, 2024). As such, when evaluating the security offered by each
tool, users should consider the risks and challenges associated with data sources used to train
the tool, as well as the features offered to protect data input into the system against leakage
(Shi, 2023).

There are many GenAI tools available (Panda & Kaur, 2024). Well-known, publicly
available and free products, such as ChatGPT (OpenAI), CoPilot (Microsoft) or Claude
(Anthropic), are made available to users according to privacy policies, terms of service and/or
use and other legalese. Most of the free tools also have a 'paid' version, which promise extra
security and privacy. Many higher education institutions subscribe to these more secure
versions, which often provide increased levels of security, ensuring a higher degree of safety
and adherence to research integrity. Researchers should investigate the appropriateness and
relevance of these secure versions for their particular research context.

Approaches currently employed by developers to address privacy and security
concerns include Privacy-Preserving Techniques (PPTs), Adversarial Defense Mechanisms,
and Regulatory Measures and Policies. These techniques serve to preserve privacy, limit the
exposure of sensitive information, and counteract malicious attacks. However, each technique
has limitations, often necessitating a trade-off between privacy and utility. As such, a compelling need persists for the development of more robust and advanced security measures for these systems (Golda et al., 2024).

Regardless of the product cost, vendor reputation, or claims in policies and terms, there remains a risk of data leakage or misuse. Developing a verification strategy to ensure no data is leaking from a system owned by a third-party provider is challenging and, consequently, it is difficult to evaluate the security guarantees offered (Gorcenski, 2023). It is possible to operate a GenAI system in a 'private' manner, for example, using open-weight large language models such as Llama (Meta). While deploying such systems has traditionally required significant technical expertise and financial resources, recent software solutions like LM Studio have simplified the process, making it easy and free of charge to run open models on a local computer. It should be noted, however, that these capabilities would be limited by the computational resources of the local hardware.

In evaluating the security risk of using any GenAI tool, researchers should consider the sensitivity of the data being entered, noting that even the most secure systems are susceptible to unintended and malicious data breaches (Chen & Esmaeilzadeh, 2024). Researchers are encouraged to investigate the use of secure and protected versions, such as those offered through institutional subscriptions and models operating on local hardware, that adequately meet ethical requirements. Extreme care should be taken to avoid disclosing any sensitive personal information, confidential commercial/organisational information and data associated with national security, noting that even anonymised patterns in data could potentially reidentify individuals if models are improperly handled after training.

## L - Look at user agreements

Finally, it is essential that researchers carefully review the user agreements, including the terms and conditions of the AI tools they are employing, to make informed decisions regarding their use. These documents often contain critical information about the ownership of inputs and outputs, which can significantly affect rights to the content created using the AI tool. User input data (known as "input") or content generated by GenAI (known as "output") raises ownership concerns. In some user agreements, the terms regarding ownership of the output are unclear. For instance, does the user or the AI company own the output? This is particularly important for commercial applications, where researchers may wish to protect AI-generated works.

Understanding the ownership of data and the resulting outputs is crucial for maintaining control over the researcher's intellectual property and ensuring their work is not exploited without their consent. Additionally, user agreements provide details on data storage, which has significant implications for privacy and security. Although the Italian Data Protection Authority emphasised that the widespread collection and storage of personal data for ChatGPT's training lacks a solid legal basis (Powell, 2023), some user agreements allow for input data to be used for model training, which adds to privacy concerns. Researchers should carefully examine these aspects of the agreements to ensure that the AI tool complies with relevant data protection regulations within their context and that their data is managed in a manner consistent with ethical research standards and legal obligations.

Some agreements give companies the right to change terms at any time, potentially leading to situations where users unknowingly agree to new conditions that alter how their data is used or who owns the generated content. For instance, OpenAI's Terms of Use include a clause stating that the company reserves the right to change its terms at any time (OpenAI, 2023). The same applies to Google, which also reserves the right to adjust its terms and services as needed (Google LLC, 2024). Users are bound by these updated terms by continuing to use the service, and while OpenAI may provide notice for some changes (such as price increases), other modifications can be implemented without direct communication. Users might not fully realise when and how the terms have changed, which could affect how their data is processed or how AI-generated outputs are owned.

In summary, it is important for researchers to carefully review the terms of service and privacy policies when using GenAI applications, as there can be implications related to data privacy and intellectual property. They should also check the terms and conditions regularly as these may change during their engagement with the GenAI tool. Researchers need to adhere to the research integrity requirements of their project and, therefore, must ensure that what they are agreeing to meets the requirements of their particular situation. This thorough scrutiny of terms and conditions helps prevent potential legal issues and enhances the researcher's ability to use GenAI tools responsibly and ethically, safeguarding their work and the sensitive information involved.

## Conclusion

The ETHICAL Framework presented in this article stands as a foundational resource for researchers navigating the ethical challenges associated with GenAI. While existing literature and guidelines outline the ethical issues researchers need to consider, this framework progresses beyond awareness to practical action. The ETHICAL Framework explicitly equips researchers with actionable principles, providing clear guidance on ethical GenAI use in research, thereby supporting both integrity and impact.

Each component of the ETHICAL Framework represents a call to action. Researchers are urged not just to consider but to actively *Examine, Think about, Harness, Indicate, Critically engage, Access*, and *Look* at aspects of GenAI use. By integrating these actions into their research workflows, researchers will foster transparency, mitigate risks, and uphold the social responsibility integral to academia. The framework emphasises that ethical GenAI use is not passive – it is a process that demands critical, ongoing engagement with each tool's functionalities and limitations, coupled with an informed understanding of legal and social implications.

The significance of this framework extends beyond research. For instance, in teaching and learning environments, the ETHICAL Framework could guide both educators and students in the ethical use of GenAI within both assessment and feedback. As educational institutions increasingly integrate AI-driven tools into the curriculum, the framework can offer structured guidance for ethical AI literacy development, supporting students, educators, and administrators to engage responsibly with technology in their teaching and learning practices. It also provides insights for regulators, research institutions and GenAI developers guiding these stakeholders to better support researchers and/or design GenAI-enhanced research tools more effectively.

However, implementing this framework brings challenges. It must be acknowledged that integrating these principles consistently may require time, resources, and institutional commitment, particularly in balancing practical application with existing institutional guidelines. There is also the potential for ethical GenAI use to become incidental if these principles are not actively championed, risking erosion of trust, inclusivity, and accountability within both research and educational contexts.

As such, the ETHICAL Framework should be seen as a guidepost, with researchers encouraged to critically reflect on its application – identifying which components may prove most challenging to implement and considering what resources are necessary to uphold ethical standards. Ultimately, by committing to this framework, researchers contribute to shaping the future of AI-enhanced scholarship in a way that reinforces ethical responsibility for future generations.

## Acknowledgement of AI use

We acknowledge that AI tools including, Microsoft Copilot, and a customised version of ChatGPT 4 (OpenAI, https://chat.openai.com/) were used in parts of this document to assist with idea generation, and refine phrasing and academic tone of the writing. Where AI outputs were used, they were adapted to reflect the authors' own styles and voices. The authors take full responsibility for the final content of this document.

## Funding

This project was partially funded by a University of Southern Queensland, Early Career Researcher Grant.